# Experimental observation of melting of the effective Minkowski spacetime in cobalt-based ferrofluids

Igor I. Smolyaninov [1], Vera N. Smolyaninova [2]

[1] *Department of Electrical and Computer Engineering, University of Maryland, College Park, MD 20742, USA*

[2] *Department of Physics Astronomy and Geosciences, Towson University,*

*8000 York Rd., Towson, MD 21252, USA*

**Hyperbolic metamaterials were originally introduced to overcome the diffraction limit of optical imaging. Soon thereafter it was realized that they demonstrate a number of novel phenomena resulting from the broadband singular behavior of their density of photonic states. These novel phenomena and applications include microscopy, stealth technologies, enhanced quantum-electrodynamic effects, thermal hyperconductivity, superconductivity, and interesting gravitation theory analogues. Here we describe the behaviour of cobalt nanoparticle-based ferrofluid in the presence of an external magnetic field, and demonstrate that it forms a self-assembled hyperbolic metamaterial, which may be described as an effective 3D Minkowski spacetime for extraordinary photons. Moreover, such photons perceive thermal gradients in the ferrofluid as analogue of gravitational field, which obeys the Newton law. If the magnetic field is not strong enough, the effective Minkowski spacetime gradually melts under the influence of thermal fluctuations. On the other hand, it may restore itself if the magnetic field is increased back to its original value. Direct microscopic visualization of such a Minkowski spacetime melting/crystallization is presented, which is somewhat similar to hypothesized formation of the Minkowski spacetime in loop quantum cosmology and may mimic various cosmological Big Bang scenarios.**

# 1. Introduction

Hyperbolic metamaterials are extremely anisotropic uniaxial materials, which behave like a metal in one direction and like a dielectric in the orthogonal direction. Originally introduced to overcome the diffraction limit of optical imaging [1-2], hyperbolic metamaterials demonstrate a number of novel phenomena resulting from the broadband singular behavior of their density of photonic states, which range from super resolution imaging [2-4] to enhanced quantum-electrodynamic effects [5,6], new stealth technology [7], thermal hyperconductivity [8], high Tc superconductivity [9], and interesting gravitation theory analogues [10]. Recent developments in gravitation theory provide numerous clues, which strongly indicate that the classic general relativity is an effective macroscopic theory, which will be eventually replaced with a more fundamental theory based on yet unknown microscopic degrees of freedom [11,12]. Unfortunately, these true microscopic degrees of freedom cannot be probed directly. Our ability to obtain experimental insights into the future microscopic theory is severely limited by low energy scales available to terrestrial physics and even to astronomical observations. In order to circumvent this problem, it is instructive to look at various examples of emergent gravity and analogue spacetimes [13], which appear in such solid state systems as superfluid helium, electromagnetic metamaterials, and cold atomic Bose-Einstein condensates. By looking at such systems we may better understand how macroscopic effective gravity may arise from the well-known microscopic atomic degrees of freedom.

Very recently magnetic ferrofluids subjected to external magnetic field have emerged as an interesting example of an electromagnetic metamaterial, which exhibits gravity-like nonlinear optical interactions, and which may be described by an emergent effective Minkowski spacetime [14-16]. Metamaterials are artificially structured materials, which are built from conventional microscopic materials in order to engineer their electromagnetic, elastic, thermal, acoustic, etc. properties in a desired way. Because of considerable 3D nanofabrication difficulties, electromagnetic metamaterials are typically confined to small sizes in two spatial dimensions. Recent demonstration that a ferrofluid subjected to strong enough magnetic field self-assembles into a "wire array" hyperbolic metamaterial (see Fig.1) [14,17] opens up many new directions in metamaterial research. Since such a metamaterial is created by 3D self-assembly, its dimensions are not limited by nanofabrication issues. Unlike other typical metamaterial systems, such a macroscopic self-assembled 3D metamaterial may also exhibit reach physics associated with topological defects [18] and phase transitions. Therefore, as was pointed out recently by Mielczarek and Bojowald [19,20], the properties of self-assembled magnetic nanoparticle-based hyperbolic metamaterials exhibit strong similarities with the properties of some microscopic quantum gravity models, such as loop quantum cosmology.

Here we will describe direct microscopic visualization of phase transitions in the cobalt nanoparticle based ferrofluids. In the presence of an external magnetic field the ferrofluid forms a self-assembled hyperbolic metamaterial, which may be described as an effective "3D Minkowski spacetime" for extraordinary photons. Moreover, it appears that the extraordinary photons propagating inside the ferrofluid perceive thermal gradients in the ferrofluid as analogue of gravitational field, which obeys the Newton law. If the magnetic field is not strong enough, this effective Minkowski spacetime gradually melts under the influence of thermal fluctuations. On the other hand, it may restore itself if the magnetic field is increased back to its original value. We will present direct microscopic visualization of such a Minkowski spacetime melting/crystallization, which is somewhat similar to hypothesized formation of the Minkowski spacetime in loop quantum cosmology. It is also interesting to note that the physical vacuum appears to exhibit hyperbolic metamaterial properties when subjected to a very strong magnetic field [21,22]. Thus, the ferrofluid in an external magnetic field provides us with an easily accessible experimental system, which lets us trace the emergence of gravity and the effective Minkowski spacetime from the well-understood microscopic degrees of freedom. In addition, the ferrofluid behaviour near the metric signature transition may mimic various cosmological Big Bang scenarios. Unlike other systems exhibiting analogue gravity behaviour, the unique feature of the ferrofluid consists in our ability to directly visualize the Minkowski spacetime formation at the microscopic level.

## 2. Electromagnetic properties of cobalt nanoparticle-based ferrofluid

Our experiments use the cobalt magnetic fluid 27-0001 from Strem Chemicals, which is composed of 10 nm cobalt nanoparticles in kerosene coated with sodium dioctylsulfosuccinate and a monolayer of LP4 fatty acid condensation polymer. The average volume fraction of cobalt nanoparticles in this ferrofluid is $p = 8.2\%$. Magnetic nanoparticles in a ferrofluid are known to form nanocolumns aligned along the external magnetic field [23]. In addition, depending on the magnetic field magnitude, solvent used, and the nanoparticle concentration, the ferrofluid undergoes phase separation into nanoparticle rich and nanoparticle poor phases [17]. The typical spatial scale of phase separation is 0.1-1 μm. Therefore, the microscopic structure of the ferrofluid may be visualized directly using an optical microscope.

Macroscopic electromagnetic properties of such a ferrofluid may be understood based on the Maxwell-Garnett approximation via the dielectric permittivities $\varepsilon_m$ and $\varepsilon_d$ of cobalt and kerosene, respectively. Volume fraction of cobalt nanoparticles aligned into nanocolumns by the external magnetic field, $\alpha(H,T)$, depends on temperature and the field magnitude. At very large magnetic fields almost all nanoparticles are aligned into nanocolumns, so that $\alpha(\infty, T) = p = 8.2\%$, and a 3D wire array hyperbolic metamaterial [17,18] is formed as shown schematically in Fig.1(a). We will assume that at smaller fields the difference $\alpha(\infty,T) - \alpha(H,T)$ describes cobalt nanoparticles, which are not aligned and distributed homogeneously inside the ferrofluid. Dielectric polarizability of these nanoparticles may be included into $\varepsilon_d$, leading to slight increase in its value. Using this model, the diagonal components of the ferrofluid permittivity may be calculated using the Maxwell-Garnett approximation as follows [24]:

$$\varepsilon_z = \varepsilon_2 = \alpha(H,T)\varepsilon_m + (1 - \alpha(H,T))\varepsilon_d \tag{1}$$

$$\varepsilon_x = \varepsilon_y = \varepsilon_1 = \frac{2\alpha(H,T)\varepsilon_m\varepsilon_d + (1 - \alpha(H,T))\varepsilon_d(\varepsilon_d + \varepsilon_m)}{(1 - \alpha(H,T))(\varepsilon_d + \varepsilon_m) + 2\alpha(H,T)\varepsilon_d} \tag{2}$$

Calculated wavelength dependencies of $\varepsilon_2$ and $\varepsilon_1$ at $\alpha(\infty,T) = p = 8.2\%$ are plotted in Fig. 1(b). These calculations are based on the optical properties of cobalt in the infrared range [25]. While $\varepsilon_1$ stays positive and almost constant, $\varepsilon_2$ changes sign to negative around $\lambda = 1\mu m$. If the volume fraction of cobalt nanoparticles varies, this change of sign occurs at some critical value $\alpha_H$:

$$\alpha > \alpha_H = \frac{\varepsilon_d}{\varepsilon_d - \varepsilon_m} \ , \tag{3}$$

so that the ferrofluid becomes a hyperbolic metamaterial [14,17]. Alignment of cobalt nanoparticle chains along the direction of external magnetic field is clearly revealed by polarization dependencies of ferrofluid transmission and by microscopic images of the ferrofluid, as shown in Fig.1(c,d). For example, Fig.1(c) shows experimentally measured transmission of the cobalt based ferrofluid at $\lambda$=1.55 μm as a function of external magnetic field and polarization angle. Zero degree polarization corresponds to E field of the electromagnetic wave perpendicular to the direction of external magnetic field. The observed strong polarization dependencies are consistent with the hyperbolic character of the ferrofluid. As far as magnetic permeability is concerned, at the visible and infrared frequencies the ferrofluid may be considered as a non-magnetic ($\mu$=1) medium.

Propagation of monochromatic extraordinary photons inside the ferrofluid is described by the wave equation, which is formally equivalent to a 3D Klein-Gordon equation for a massive field $\varphi_\omega = E_z$ in a 3D Minkowski spacetime:

$$-\frac{\partial^2 \varphi_\omega}{\varepsilon_1 \partial z^2} + \frac{1}{(-\varepsilon_2)}\left(\frac{\partial^2 \varphi_\omega}{\partial x^2} + \frac{\partial^2 \varphi_\omega}{\partial y^2}\right) = \frac{\omega_0^2}{c^2}\varphi_\omega = \frac{m^{*2}c^2}{\hbar^2}\varphi_\omega \qquad (4)$$

in which the spatial coordinate $z$ behaves as a timelike variable, and the effective mass is $m^* = \hbar\omega_0/c^2$ [26,27]. Eq.(4) exhibits effective Lorentz invariance under the coordinate transformation

$$z' = \frac{1}{\sqrt{1 - \frac{\varepsilon_1}{(-\varepsilon_2)}\beta}}(z - \beta x) \qquad (5)$$

$$x' = \frac{1}{\sqrt{1 - \frac{\varepsilon_1}{(-\varepsilon_2)}\beta}}\left(x - \beta\frac{\varepsilon_1}{(-\varepsilon_2)}z\right),$$

where $\beta$ is the effective boost. Similar to our own Minkowski spacetime, the effective Lorentz transformations in the $xz$ and $yz$ planes form the Poincare group together with translations along $x$, $y$, and $z$ axis, and rotations in the $xy$ plane. Thus, the wave equation (4) describes world lines of massive particles which propagate in an effective 2+1 dimensional Minkowski spacetime. Components of the metamaterial dielectric tensor define the effective metric $g_{ik}$ of this spacetime: $g_{00}=-\varepsilon_1$ and $g_{11}=g_{22}=-\varepsilon_2$. Nonlinear optical Kerr effect bends this spacetime resulting in effective gravitational interaction between extraordinary photons [15,16]. When the ferrofluid develops phase separation into cobalt rich and cobalt poor phases, its microscopic structure and the local field intensity in the hyperbolic frequency range may be observed using an infrared microscope. As a result, microscopic properties of the effective Minkowski spacetime, as well as its melting/crystallization may be directly visualized in the experiment. In order to better understand these experiments and their connection with gravity theory, let us consider in more detail the emergence of analogue gravity in ferrofluids.

## 3. Analogue gravity in ferrofluids

When the nonlinear optical effects become important, they are described in terms of various order nonlinear susceptibilities $\chi^{(n)}$ of the metamaterial:

$$D_i = \chi_{ij}^{(1)} E_j + \chi_{ijl}^{(2)} E_j E_l + \chi_{ijlm}^{(3)} E_j E_l E_m + ... \qquad (6)$$

Taking into account these nonlinear terms, the dielectric tensor of the metamaterial (which defines its effective metric) may be written as

$$\varepsilon_{ij} = \chi_{ij}^{(1)} + \chi_{ijl}^{(2)} E_l + \chi_{ijlm}^{(3)} E_l E_m + ... \qquad (7)$$

In a centrosymmetric material all the second order nonlinear susceptibilities $\chi_{ijl}^{(2)}$ must be equal to zero. It is clear that eq.(7) provides coupling between the matter content (photons) and the effective metric of the metamaterial spacetime. Nonlinear optical effects "bend" this effective Minkowski spacetime, resulting in gravity-like interaction of extraordinary light rays. It appears that in the weak field limit the nonlinear optics of hyperbolic metamaterials may indeed be formulated as an effective gravity [15]. In such a limit the Einstein equation

$$R_i^k = \frac{8\pi\gamma}{c^4} \left( T_i^k - \frac{1}{2} \delta_i^k T \right) \qquad (8)$$

is reduced to

$$R_{00} = \frac{1}{c^2}\Delta\phi = \frac{1}{2}\Delta g_{00} = \frac{8\pi\gamma}{c^4}T_{00} \quad , \qquad (9)$$

where $\phi$ is the gravitational potential [28]. Therefore, the third order terms in eq.(7) may provide correct coupling between the effective metric and the energy-momentum tensor. These terms are associated with the optical Kerr effect. The detailed analysis in [15] indeed indicates that the Kerr effect in a hyperbolic metamaterial leads to effective gravity.

Let us briefly reproduce this analysis, since it is important for the discussion of connection between the effective gravity and the thermodynamics of a magnetized ferrofluid. Since $z$ coordinate plays the role of time, while $g_{00}$ is identified with $-\varepsilon_1$, eq.(9) may be re-written as

$$-\Delta^{(2)}\varepsilon_1 = \frac{16\pi\gamma*}{c^4}T_{zz} = \frac{16\pi\gamma*}{c^4}\sigma_{zz} \quad , \qquad (10)$$

where $\Delta^{(2)}$ is the 2D Laplacian operating in the $xy$ plane, $\gamma*$ is the effective "gravitation constant", and $\sigma_{zz}$ is the $zz$ component of the Maxwell stress tensor of the electromagnetic field in the medium:

$$\sigma_{zz} = \frac{1}{4\pi}\left(D_z E_z + H_z B_z - \frac{1}{2}\left(\vec{D}\vec{E} + \vec{H}\vec{B}\right)\right) \qquad (11)$$

Taking into account eq.(4), for a single plane wave eq.(10) may be rewritten as [15]

$$-\Delta^{(2)}\varepsilon_1 = -\Delta^{(2)}\left(\varepsilon_1^{(0)} + \delta\varepsilon_1\right) = k_x^2\delta\varepsilon_1 = -\frac{4\gamma*B^2 k_z^2}{c^2\omega^2\varepsilon_1} \quad , \qquad (12)$$

where without a loss of generality we have assumed that the $B$ field of the wave is oriented along $y$ direction. We also assumed that the nonlinear corrections to $\varepsilon_1$ are small, so that we can separate $\varepsilon_1$ into the constant background value $\varepsilon_1^{(0)}$ and the weak nonlinear corrections. These nonlinear corrections do indeed look like the Kerr effect assuming that the extraordinary photon wave vector components are large compared to $\omega/c$:

$$\delta\varepsilon_1 = -\frac{4\gamma * B^2 k_z^2}{c^2 \omega^2 \varepsilon_1 k_x^2} \approx \frac{4\gamma * B^2}{c^2 \omega^2 \varepsilon_2} = \chi^{(3)} B^2 \qquad (13)$$

This assumption has to be the case if extraordinary photons may be considered as classic "particles". Unlike the usual "elliptic" optical materials, this assumption is justified by the hyperbolic dispersion relation of the extraordinary photons:

$$\frac{\omega^2}{c^2} = \frac{k_z^2}{\varepsilon_1} - \frac{k_x^2 + k_y^2}{|\varepsilon_2|} \qquad (14)$$

which follows from eq.(4). Eq.(13) establishes connection between the effective gravitation constant $\gamma*$ and the third order nonlinear susceptibility $\chi^{(3)}$ of the hyperbolic metamaterial. Since $\varepsilon_2 < 0$, the sign of $\chi^{(3)}$ must be negative for the effective gravity to be attractive. This condition is satisfied naturally in most liquids, and in particular, in kerosene. Because of the large and negative thermo-optic coefficient inherent to most liquids, heating produced by partial absorption of the propagating beam translates into a significant decrease of the refractive index at higher light intensity. For example, the reported thermo-optic coefficient of water reaches $\Delta n/\Delta T = -5.7 \times 10^{-4} K^{-1}$ [29]. Moreover, introduction of nanoparticles or absorbent dye into the liquid allows for further increase of the thermal nonlinear response [30]. Therefore, a ferrofluid-based self-assembled hyperbolic metamaterial naturally exhibits effective gravity. The thermal origin of this effective gravity looks interesting in light of the modern advances in gravitation theory [11,12], which strongly indicate that the classic general relativity description of gravity results from thermodynamic effects.

As a next step, let us consider the effect of thermal gradients in the ferrofluid on its effective metric. It appears that the extraordinary photons propagating inside the ferrofluid perceive thermal gradients as an effective gravitational field. Indeed, eqs. (1,2) imply that $\varepsilon_2$ and $\varepsilon_1$ (which may be understood as the effective metric coefficients $g_{00}=-\varepsilon_1$ and $g_{11}=g_{22}=-\varepsilon_2$ of the metamaterial spacetime) depend on the volume fraction $\alpha(H,T)$ of cobalt nanoparticles aligned into nanocolumns by the external magnetic field, which in turn depends on the local temperature of the ferrofluid. Since ferrofluids subjected to external magnetic field are known to exhibit classical superparamagnetic behaviour [31], well established results from the theory of magnetism may be used to calculate $\alpha(H,T)$. Superparamagnetism is a form of magnetism that is exhibited by magnetic materials, which consist of small ferromagnetic or ferrimagnetic nanoparticles. Superparamagnetism occurs in nanoparticles which are single-domain, which is possible when their diameter is below ~50 nm, depending on the material. Since the typical size of magnetic nanoparticles in ferrofluids is ~ 10 nm, magnetic ferrofluids also belong to the class of superparamagnetic materials. When an external magnetic field is applied to an assembly of superparamagnetic nanoparticles, their magnetic moments tend to align along the applied field, leading to a net magnetization. If all the particles may be considered to be roughly identical (as in the case of a homogeneous ferrofluid), and the temperature is low enough, then the magnetization of the assembly is [31]

$$M(H,T) = n\mu \tanh\left(\frac{\mu H}{kT}\right) \quad , \qquad (15)$$

where $n$ is the nanoparticle concentration, and $\mu$ is their magnetic moment. Therefore, within the scope of our model of the dielectric response of the ferrofluid (see eqs.(1,2)), we may assume that

$$\alpha(H,T) = \alpha_\infty \tanh\left(\frac{\mu H}{kT}\right) \qquad (16)$$

where $\alpha_\infty = 0.082$. Let us consider the typical case of $-\varepsilon_m >> \varepsilon_d$ and assume that $\alpha(H,T)$ is small (which is typically required for the Maxwell-Garnett approximation to be valid). In such a case the effective metric coefficients are:

$$g_{11} = g_{22} = -\varepsilon_2 \approx -\alpha_\infty \tanh\left(\frac{\mu H}{kT}\right)\varepsilon_m - \varepsilon_d \qquad (17)$$

$$g_{00} = -\varepsilon_1 \approx -\varepsilon_d\left(1 + 2\alpha_\infty \tanh\left(\frac{\mu H}{kT}\right)\right) \qquad (18)$$

The effective spacetime appears to be a Minkowski one if the temperature is low enough, so that

$$\alpha_\infty \tanh\left(\frac{\mu H}{kT}\right) > \frac{\varepsilon_d}{\left(-\varepsilon_m\right)} \qquad (19)$$

Thus, in the weak field limit the effective gravitational potential (see eq.(9)) is

$$\phi = \frac{c^2}{2}\left(\frac{g_{00}}{\varepsilon_d} - 1\right) = \alpha_\infty c^2\left(\tanh\left(\frac{\mu H}{kT}\right) - 1\right) \qquad (20)$$

(where the value of the effective potential at $T$=0 is chosen as a reference) and the effective gravitational field $\vec{G}$ is

$$\vec{G} = -\nabla\phi = \frac{\alpha_\infty c^2 \mu H}{kT^2 \cosh^2\left(\frac{\mu H}{kT}\right)}\nabla T \qquad (21)$$

where we have assumed the external magnetic field $H$ to be constant. Thus, the extraordinary photons propagating inside the ferrofluid perceive thermal gradients in the ferrofluid as an effective gravitational field. This observation correlates nicely with the thermal origin of gravity-like nonlinear optical interaction of the extraordinary light rays due to photo-thermal Kerr effect, which was noted above.

Let us consider a linear source of heat $q$ (see Fig.2) placed inside the ferrofluid, which is constant in time and aligned parallel to the external magnetic field $H$ (and therefore it is also parallel to the nanoparticle chains). Let us also assume that the ferrofluid is kept at a constant temperature $T_0$. The temperature distribution around the source is defined by the two-dimensional heat conductance equation

$$-\sigma\left(\frac{\partial^2}{\partial x^2}+\frac{\partial^2}{\partial y^2}\right)T = -\sigma\nabla(\nabla T) = \frac{1}{c_p\rho}q \qquad (22)$$

where $\sigma$ is the thermal diffusivity, $c_p$ is the heat capacity and $\rho$ is the density of the ferrofluid. If the linear source of heat is weak enough, so that the temperature-dependent terms in the denominator of eq.(21) may be considered constant, eq.(22) may be re-written as an equation for the effective gravitational field $\vec{G}$ :

$$\nabla\vec{G} = -\frac{\alpha_\infty c^2\mu H}{\sigma c_p\rho k T_0^2\cosh^2\left(\dfrac{\mu H}{kT_0}\right)}q = -\gamma^* q \qquad (23)$$

which has the form of the Newton law of gravity. As evident from eq.(23), the heat source $q$ plays the role of a gravitational mass. Thus, a linear source of heat parallel to the external magnetic field $H$ behaves as a world line of massive gravitating object in the metamaterial Minkowski spacetime. It is interesting that unlike quantum mechanical derivation based on the holographic principle reported in [12], the Newton law in a ferrofluid arises in a purely classical Boltzmann system. In principle, a similar approach may be applied to other magnetic systems at lower temperatures, where the thermodynamics description becomes explicitly quantum mechanical [32]. However, direct visualization of gravity-like effects in such systems appears to be much more difficult.

As far as the strong field limit is concerned (where the temperature may not be considered almost constant) it was pointed out in ref.[15] that at large enough power gravitational self-interaction of the extraordinary rays is strong enough, so that spatial solitons may be formed. These spatial solitons also behave as world lines of compact self-gravitating bodies in the effective 2+1 dimensional Minkowski spacetime. This is not surprising since due to photo-thermal effect solitons may also be considered as linear sources of heat. Moreover, upon increase of the optical power, a spatial soliton may collapse into a black hole analogue [15]. Indeed, in the presence of self-defocusing negative Kerr effect in the dielectric host $\varepsilon_d$, the wave equation (eq.(4)) must be modified. Assuming a spatial soliton-like solution which conserves energy per unit length $W \sim P/c$ (where $P$ is the laser power), the soliton width $\rho$ and the magnetic field amplitude $B$ of the extraordinary wave are related as

$$B^2 \rho^2 = P/c \qquad (24)$$

As a result, eq.(4) must be re-written as

$$-\frac{\partial^2 \varphi_\omega}{\left(\varepsilon_1^{(0)} - \frac{\left(-\chi^{(3)}\right)P}{c\rho^2}\right)\partial z^2} + \frac{1}{\left(-\varepsilon_2\right)}\left(\frac{\partial^2 \varphi_\omega}{\partial x^2} + \frac{\partial^2 \varphi_\omega}{\partial y^2}\right) = \frac{\omega_0^2}{c^2}\varphi_\omega \qquad (25)$$

where $\varepsilon_l^{(0)}$ is the dielectric permittivity component at $P$=0 (note that the nonlinear contribution to $\varepsilon_2 \approx \alpha\varepsilon_m$ may be neglected). Effective metric described by eq.(25) has a black hole-like singularity at

$$\rho = \left(\frac{\left(-\chi^{(3)}\right)P}{c\varepsilon_1^{(0)}}\right)^{1/2} \qquad (26)$$

The critical soliton radius at $P$=100W was estimated to be ~20 nm [15], which does not look completely unrealistic from the fabrication standpoint. While achieving such critical values with CW lasers seems implausible, using pulsed laser definitely looks like a realistic option since thermal damage produced by a pulsed laser typically depends on the pulse energy and not the pulse power.

## 4. Microscopic investigation of Minkowski spacetime melting

Another "strong field" situation corresponds to the ferrofluid behavior near the metric signature transition. It appears that "melting" of the effective Minkowski spacetime in such a case may mimic various cosmological Big Bang scenarios. This is not surprising since symmetry breaking in magnetic systems is typically described by the Mexican hat potential. Moreover, such a Big Bang-like behavior may be observed directly using an optical microscope. Indeed, based on eq.(4) and Fig.1(b) it is clear that the factor $(-\varepsilon_2)$ plays the role of a scale factor of the effective Minkowski spacetime

$$ds^2 = -\varepsilon_1 dz^2 + (-\varepsilon_2)\left(dx^2 + dy^2\right) \qquad (27)$$

(since according to eq.(18) $\varepsilon_1$ is positive and almost constant, as plotted in Fig.1(b)). The scale factor of the effective Minkowski spacetime calculated using eq.(17) is plotted in Fig.3 as a function of $kT/\mu H$ at different values of $-\alpha_\infty \varepsilon_m$. Let us assume that the temperature distribution inside the ferrofluid may be described as

$$T = T_c - z\nabla T = T_c - \eta z \ , \qquad (28)$$

where $T_c$ is the temperature of the metric signature transition (the $\varepsilon_2$=0 point). Since z coordinate plays the role of time in the effective spacetime described by eq.(27), such a temperature gradient $\eta$ will result in a Big Bang-like spacetime expansion described by the scale factor $-\varepsilon_2(T_c - \eta\, z)$ plotted in Fig.3(a). Indeed, as the ferrofluid temperature falls away from the $T_c$ boundary at z=0, the spacetime scale factor increases sharply as a function of z. Note that at larger values of $-\alpha_\infty \varepsilon_m$ expansion of the effective spacetime accelerates at lower temperatures. As demonstrated below, this "cosmological" spacetime expansion may be visualized directly using an optical microscope.

We should also note that in the case of non-constant z-dependent $\varepsilon_1 = \varepsilon_x = \varepsilon_y$ and $\varepsilon_2 = \varepsilon_z$ the electromagnetic field separation into the ordinary and the extraordinary components remains well defined [33,34]. Taking into account z derivatives of $\varepsilon_1$ and $\varepsilon_2$, eq.(4) becomes

$$-\frac{\partial^2 \varphi_\omega}{\varepsilon_1 \partial z^2} + \frac{1}{(-\varepsilon_2)}\left(\frac{\partial^2 \varphi_\omega}{\partial x^2} + \frac{\partial^2 \varphi_\omega}{\partial y^2}\right) + \left(\frac{1}{\varepsilon_1^2}\left(\frac{\partial \varepsilon_1}{\partial z}\right) - \frac{2}{\varepsilon_1 \varepsilon_2}\left(\frac{\partial \varepsilon_2}{\partial z}\right)\right)\left(\frac{\partial \varphi_\omega}{\partial z}\right) +$$
$$+ \frac{\varphi_\omega}{\varepsilon_1 \varepsilon_2}\left(\frac{1}{\varepsilon_1}\left(\frac{\partial \varepsilon_1}{\partial z}\right)\left(\frac{\partial \varepsilon_2}{\partial z}\right) - \left(\frac{\partial^2 \varepsilon_2}{\partial z^2}\right)\right) = \frac{\omega_0^2}{c^2}\varphi_\omega \qquad (29)$$

Since $\varepsilon_1$ remains almost constant in a very broad temperature range far from $T$=0 (as evident from Fig.3(b)) its derivatives may be neglected. If we also neglect the second derivative of $\varepsilon_2$, the wave equation for the extraordinary field $\varphi_\omega = E_z$ may be re-written as

$$-\frac{\partial^2 \varphi_\omega}{\varepsilon_1 \partial z^2} + \frac{1}{(-\varepsilon_2)}\left(\frac{\partial^2 \varphi_\omega}{\partial x^2} + \frac{\partial^2 \varphi_\omega}{\partial y^2}\right) - \frac{2}{\varepsilon_1 \varepsilon_2}\left(\frac{\partial \varepsilon_2}{\partial z}\right)\left(\frac{\partial \varphi_\omega}{\partial z}\right) = \frac{\omega_0^2}{c^2}\varphi_\omega \qquad (30)$$

It is easy to verify that the latter equation coincides with the Klein-Gordon equation [35] for a massive particle in a gravitational field described by the metric coefficients $g_{00} = -\varepsilon_1$ and $g_{11} = g_{22} = -\varepsilon_2$:

$$\frac{1}{\sqrt{-g}} \frac{\partial}{\partial x^i} \left( g^{ik} \sqrt{-g} \frac{\partial \psi}{\partial x^k} \right) = \frac{m^2 c^2}{\hbar^2} \psi \tag{31}$$

Indeed, for a wave field $\psi = \left(-\varepsilon_2\right)^{1/2} \phi$ eq.(30) is reproduced if we also neglect the second derivative terms.

As demonstrated by Fig.1(d), the alignment of cobalt nanoparticles along the timelike z direction is easy to visualize. Our experiments were performed using a 10 μm pathlength optical cuvette placed under the infrared microscope. The cuvette was filled with the ferrofluid and illuminated from below with a linear polarized 1.55 μm laser light. The microscopic transmission images and movies were studied as a function of magnitude and direction of the external magnetic field as it was repeatedly turned on and off. Examples of the experimentally measured microscopic transmission images of the cobalt nanoparticle-based ferrofluid obtained at 30 frames per second while the external magnetic field was gradually increased from zero to about 300 G are shown in Fig.4. The frame number is indicated in each image. These images correspond to gradual "crystallization" of the effective Minkowski spacetime starting from the initial state, which may be described as an "Euclidean optical space" (since the initial state corresponds to a homogeneous isotropic ferrofluid having $\varepsilon_x = \varepsilon_y = \varepsilon_z > 0$). The pronounced periodic pattern of self-assembled stripes visible in frame #295 and #315 appears due to phase separation. The periodic stripes are oriented along the direction of the magnetic field, leading to the hyperbolic character ($\varepsilon_2 < 0$ and $\varepsilon_1 > 0$) of the ferrofluid. It is interesting to note that a sudden spike in transmission is observed around frame #235, well in advance of formation of the periodic long range order, which becomes visible after frame #275. The temporal dependence of this short spike is presented in Fig.5, which shows temporal behaviour of the average light intensity integrated over the frame, while the external magnetic field was gradually increased. This spike may be attributed to short range ordering of the cobalt nanochains along the direction of the external magnetic field. Since the individual nanochain size falls well below the diffraction limit, they are not visible in the optical images. The similarly sudden decrease in transmission indicates that the mutual alignment of this short range ordered nanochains is short lived.

Fig.6 presents a similar set of microscopic frames recorded while the external magnetic field was decreased from about 300 G to zero. If the external magnetic field is not strong enough, the effective Minkowski spacetime gradually melts under the influence of thermal fluctuations, as illustrated in Fig.6. Compared to ref. [34], such a melting represents a different "end of time" scenario, since the spacetime metric signature changes, and the effective Minkowski spacetime is completely replaced by the Euclidean optical space. Fourier analysis of the frames from Fig.6, which is shown in Fig.7 confirms gradual disappearance of the long range order. Gradual melting of the long range order is also clearly visible in Fig. 8, which shows magnified images of the same corner area from frames #564 and #580. These images reveal fine details of the effective Minkowski spacetime melting. The top row shows these images in the usual greyscale format, while the quasi-3D representation of the same images in the bottom row provides better visualization of the actual nanoparticle filaments. The quasi-3D frame #580 in Fig.8 reveals that instead of being aligned, the nanocolumns start to bend and tend to form closed loops upon gradual melting of their long range order. This loop formation is somewhat similar to the hypothesized asymptotic silence [19,20] in loop quantum gravity, which can be seen as a process in which continuous space decouples into a set of not interacting points. A more detailed set of magnified quasi-3D images of the ferrofluid obtained upon the long range order melting is shown in Fig.9. The frame number is indicated in each image. It is interesting to note that a small region in frame #600 apparently remains in a microscopic "Minkowski spacetime state" (while the rest of the original spacetime has already melted). It is highlighted by a yellow circle in the frame. Existence of such microscopic Minkowski spacetime regions inside the ferrofluid below the metric signature change threshold was hypothesized in [14].

It is also interesting to note that similar to [33], gradual melting of the effective Minkowski spacetime within a single frame may be considered as a toy model of inflation. Indeed, based on eq.(27) and Fig.1(b) it is clear that the factor $(-\varepsilon_2)$ plays the role of a scale factor of the effective Minkowski spacetime (since $\varepsilon_1$ is positive and almost constant). Fig.10 presents an example of such an inflation-like behaviour of the effective Minkwoski spacetime in frame #590. As demonstrated above, visibility of the filaments in ferrofluid transmission images, which may be assessed quantitatively by Fourier analysis, provides measurements of the relative value of $\alpha(H,T)$ compared to $\alpha_\infty$. A natural way to study ferrofluid behaviour near the metric signature transition is to gradually reduce the external magnetic field $H$ until the periodic alignment of cobalt nanoparticles begins to disappear. Since thermal and/or magnetic field gradients are unavoidable in such an experiment, a ferrofluid region showing gradual disappearance of filament periodicity across the image is relatively easy to find. A microscopic transmission image of such a region is presented in Fig.10(b). The filament periodicity (compare Fig.10(a) and 10(b)) gradually disappears towards the top of the image, which is verified by Fourier analysis in Fig.10(c) (note that the contrast in these Fourier images is internal). The measured dependence of the spacetime scale factor $-\varepsilon_2$ on the effective time calculated at $\lambda$=1.4 μm is shown in Fig.10(d). This wavelength is close enough to the $\lambda=1.5$ µm illumination wavelength used in the experiment, while the experimental data for the electromagnetic properties of cobalt [25] may also be used. Calculated values of $-\varepsilon_2$ are based on the Fourier analysis of smaller regions in Fig.10(b) (shown by the yellow boxes). The measured behaviour of the scale factor in this experiment corresponds to small values of $-\alpha_\infty \varepsilon_m$, which is expected at $\lambda=1.4$ µm based on the optical properties of cobalt [25]. As evident from Fig.10(d), the measured data show good agreement with theoretical calculations based on eq.(17).

## 5. Final remarks

Unfortunately, the described analogy between the extraordinary light propagation inside the ferrofluid and the dynamics of massive particles in Minkowski spacetime is in no way perfect. The main difficulty comes from the cross-coupling between extraordinary and ordinary light inside the ferrofluid, which may be caused by internal defects and strong gradients of the dielectric permittivity tensor components. Since ordinary light "lives" in Euclidean space, such a cross-coupling breaks the effective Lorentz symmetry (5) of the system. In addition, such a model is necessarily limited to 2+1 spacetime dimensions. Nevertheless, these limitations notwithstanding, it is interesting to trace the appearance of macroscopic "gravity" (even though imperfect) to thermodynamics of well understood microscopic degrees of freedom of the ferrofluid.

In conclusion, we have presented direct microscopic visualization of melting and crystallization of the effective Minkowski spacetime, which describes extraordinary light propagation in cobalt ferrofluid-based self-assembled hyperbolic metamaterials. As was pointed out recently by Mielczarek and Bojowald [19,20], the properties of such self-assembled hyperbolic metamaterials exhibit strong similarities with the properties of such microscopic quantum gravity models as loop quantum cosmology. It appears that the effective Minkowski spacetime behaviour near the metric signature transition may mimic various cosmological Big Bang scenarios, which may be observed directly using an optical microscope. Even though the considered model is applicable only to the extraordinary photons, it is quite interesting since it permits both direct theoretical analysis at the level of well-defined microscopic degrees of freedom, and it enables direct visualization of these microscopic degrees of freedom using standard optical tools.

**Acknowledgments**

This work is supported by the NSF grant DMR-1104676.

**Figure Captions**

**Fig. 1** (a) Schematic geometry of the metal nanowire-based hyperbolic metamaterial. (b) Wavelength dependencies of the real parts of $\varepsilon_z = \varepsilon_2$ and $\varepsilon_{x,y} = \varepsilon_1$ for a cobalt nanoparticle-based ferrofluid at $\alpha=8.2\%$ volume fraction of nanoparticles. While $\varepsilon_{xy}$ stays positive and almost constant, $\varepsilon_z$ changes sign to negative around $\lambda=1\mu m$. (c) Experimentally measured transmission of the cobalt based ferrofluid at $\lambda=1.55\ \mu m$ as a function of external magnetic field and polarization angle. Zero degree polarization corresponds to E field of the electromagnetic wave perpendicular to the direction of external magnetic field. The observed strong polarization dependencies are consistent with the hyperbolic character of the ferrofluid. (d) Microscopic image of cobalt nanoparticle-based ferrofluid in external magnetic field reveals nanoparticle alignment along the field direction.

**Fig. 2**. Schematic geometry of the Newton law derivation for effective gravity in a ferrofluid.

**Fig. 3**. (a) Scale factor of the effective Minkowski spacetime $g_{11}=g_{22}=-\varepsilon_2$ (calculated using eq.(17)) plotted as a function of $kT/\mu H$ at different values of $-\alpha_\infty \varepsilon_m$. It is assumed that $\varepsilon_d=2$. (b) Metric coefficient $g_{00}=-\varepsilon_1$ of the effective Minkowski spacetime calculated using eq.(18).

**Fig. 4**. Experimentally measured microscopic transmission images of cobalt nanoparticle-based ferrofluid obtained at 30 frames per second while the external magnetic field was gradually increased. An infrared microscope was used in combination with a 1.55 μm laser as an illumination source. The frame number is indicated in each image. These images correspond to gradual "crystallization" of the effective Minkowski spacetime.

**Fig. 5.** Average light intensity integrated over the frame while the external magnetic field was gradually increased plotted as a function of the frame number. A sudden spike in transmission is detected around frame #235, well in advance of formation of the long range order visible after frame #275. This spike is attributed to formation of short range order due to external magnetic field.

**Fig. 6.** Transmission images of cobalt nanoparticle-based ferrofluid obtained at 30 frames per second while the external magnetic field was gradually decreased. The frame number is indicated in each image. These images correspond to gradual "melting" of the effective Minkowski spacetime.

**Fig. 7.** Fourier analysis of the frames from Fig.4 confirms gradual disappearance of the long range order.

**Fig. 8.** Magnified images of the same area in frames #564 and #580 from Fig.4 reveal fine details of the effective Minkowski spacetime melting. Top row shows these images in the usual greyscale format, while the quasi-3D representation of the same images in the bottom row provides better visualization of the actual nanoparticle filaments.

**Fig. 9.** Magnified quasi-3D images of the effective Minkowski spacetime melting. The frame number is indicated in each image. A small region in frame 600, which remains in a microscopic "Minkowski spacetime state" (while the rest of the original spacetime has already melted) is highlighted by yellow circle. Existence of such miscroscopic Minkowski spacetime regions was hypothesized in [14].

**Fig. 10**. Analysis of gradual melting of the effective Minkowski spacetime within a single frame: (a) Microscopic transmission image of the cobalt nanoparticle-based ferrofluid taken in large external magnetic field $H$ using illumination with $\lambda=1.5$ $\mu$m light. (b) Analysis of gradual melting of the effective Minkowski spacetime within a single microscopic image of the ferrofluid (frame #590). The effective time direction is indicated by the red arrow. The effective spacetime expansion is shown schematically by the yellow cone. (d) Measured dependence of the spacetime scale factor $-\varepsilon_2$ on the effective time calculated at $\lambda=1.4$ μm. These calculations are based on the Fourier analysis (c) of smaller regions of Fig.10(b) shown by the yellow boxes. The measured data are compared with theoretical calculations based on eq.(17).

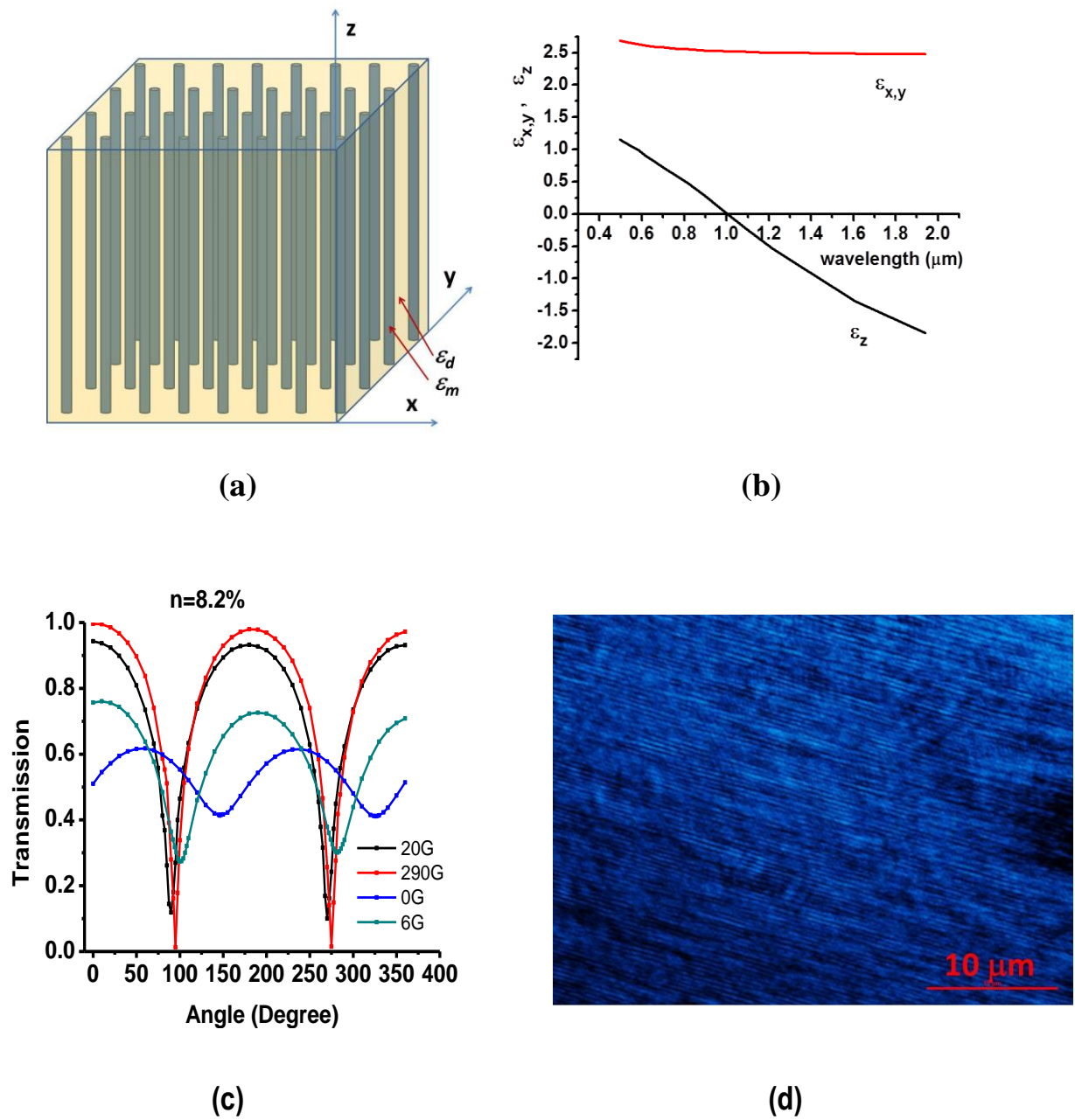

**(a)**

**(b)**

**(c)**

**(d)**

**Fig.1**

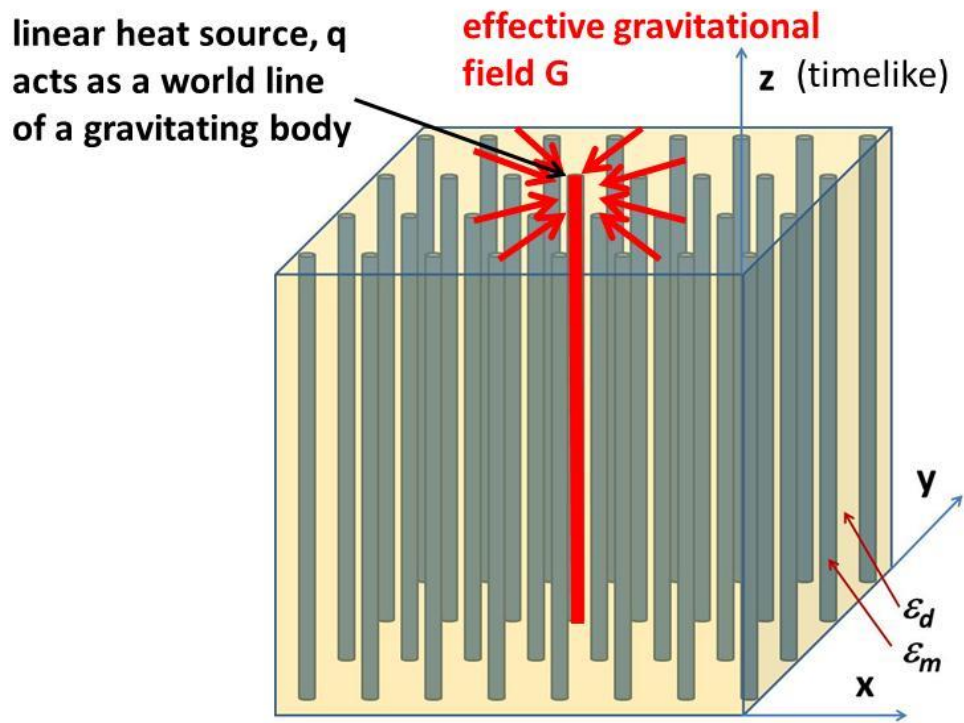

**linear heat source, q acts as a world line of a gravitating body**

**effective gravitational field G**

z (timelike)

y

x

$\varepsilon_d$
$\varepsilon_m$

**Fig. 2**

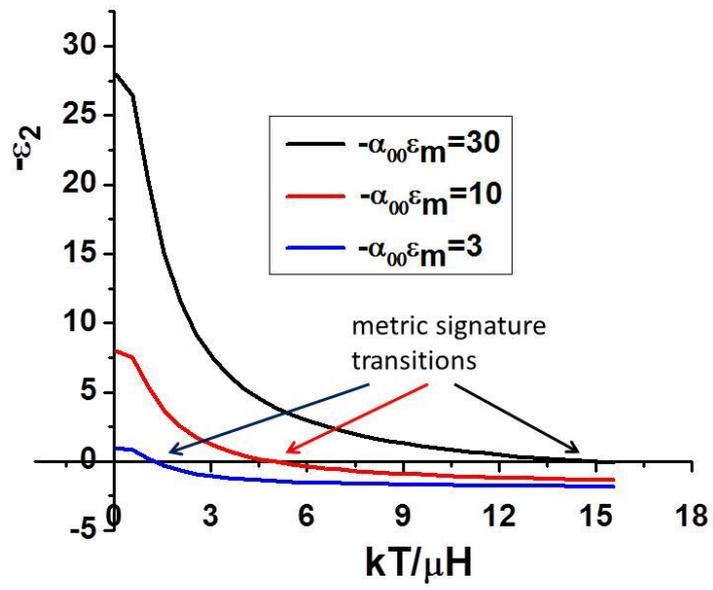

(a)

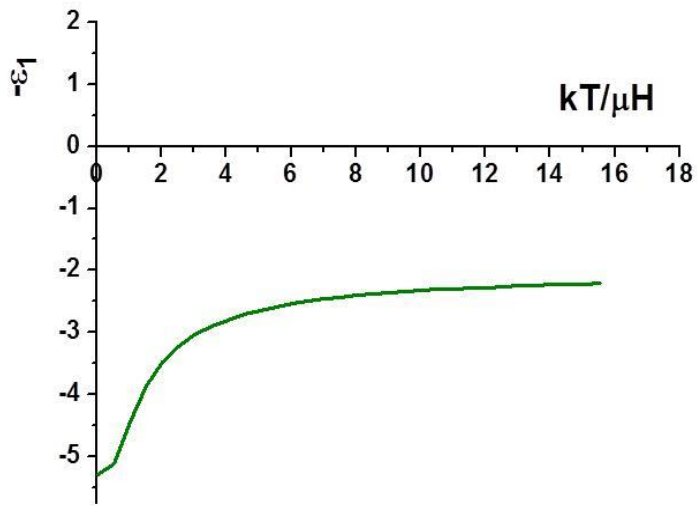

(b)

**Fig. 3**

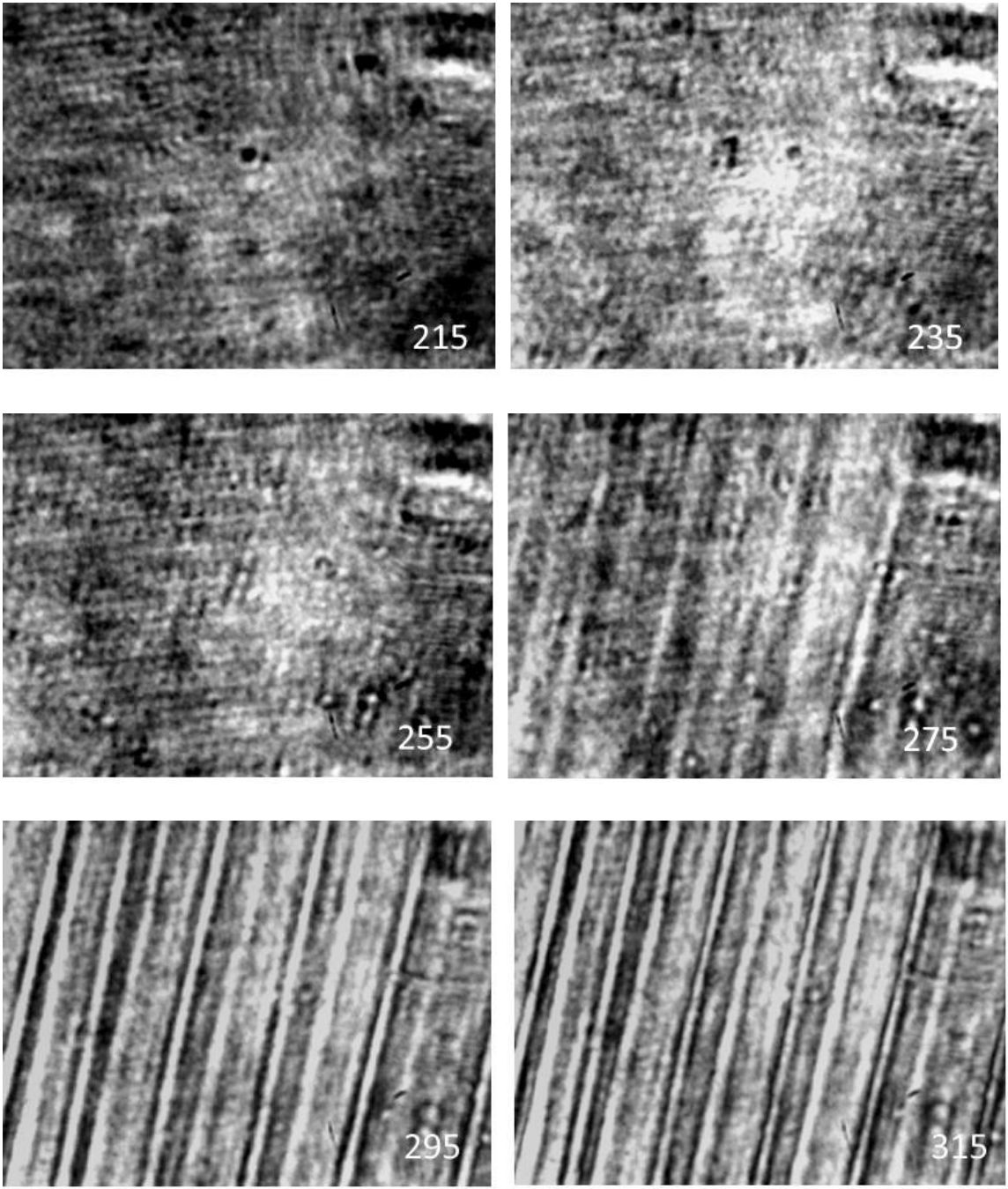

**Fig. 4**

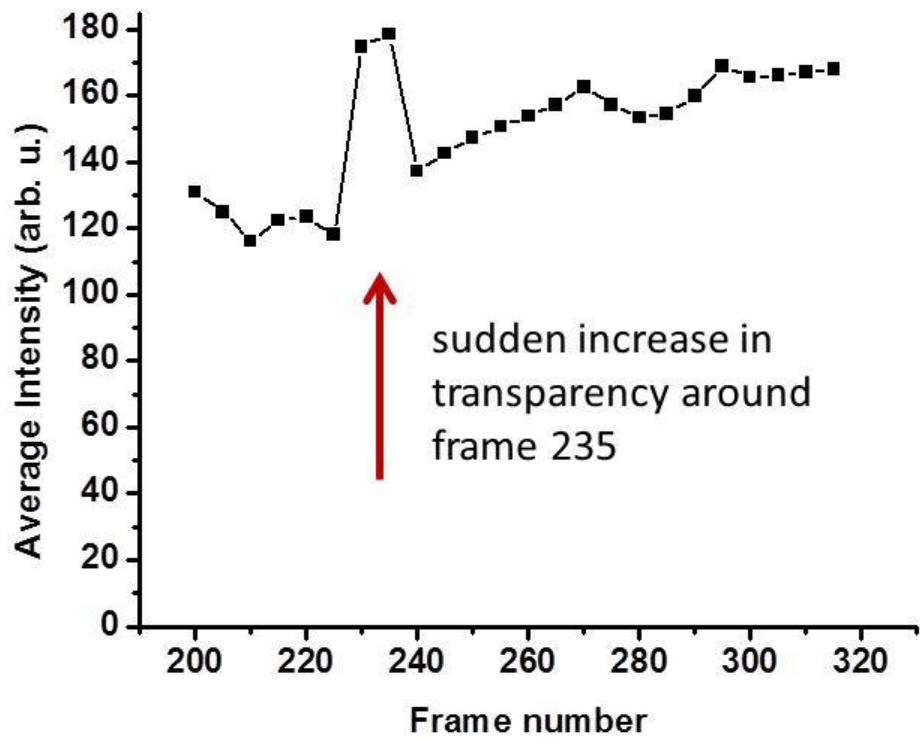

**Fig. 5**

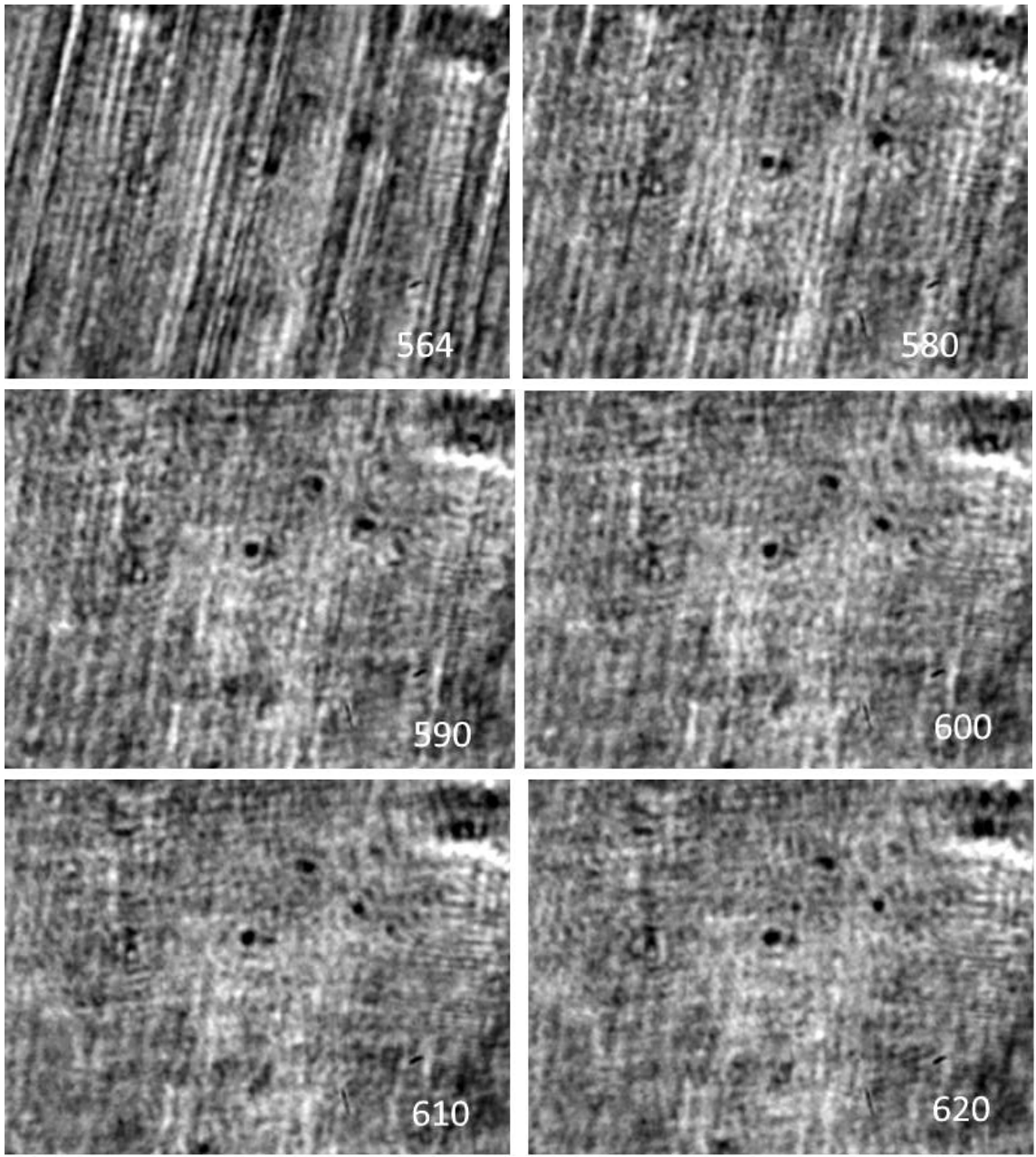

**Fig. 6**

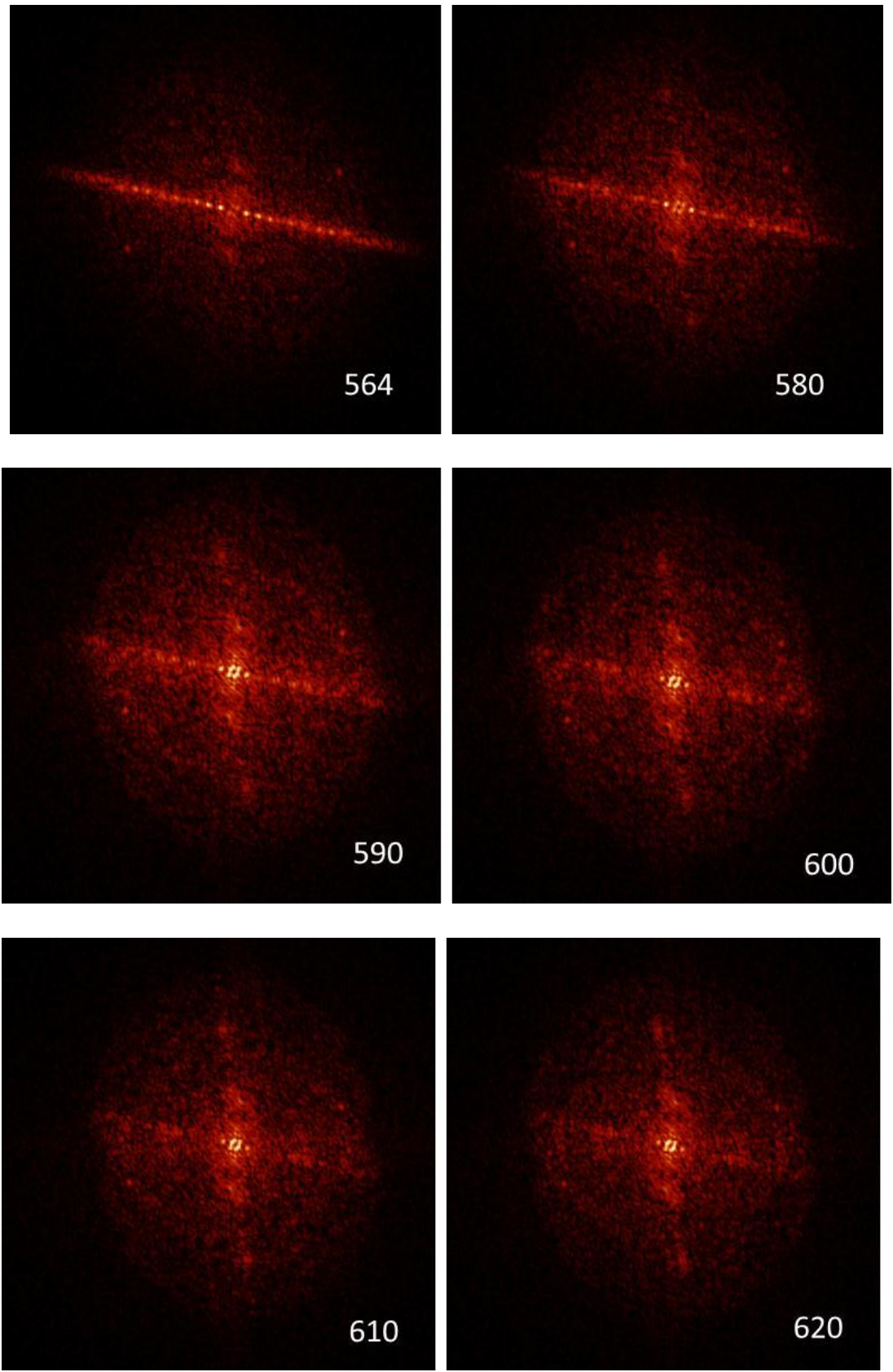

**Fig. 7**

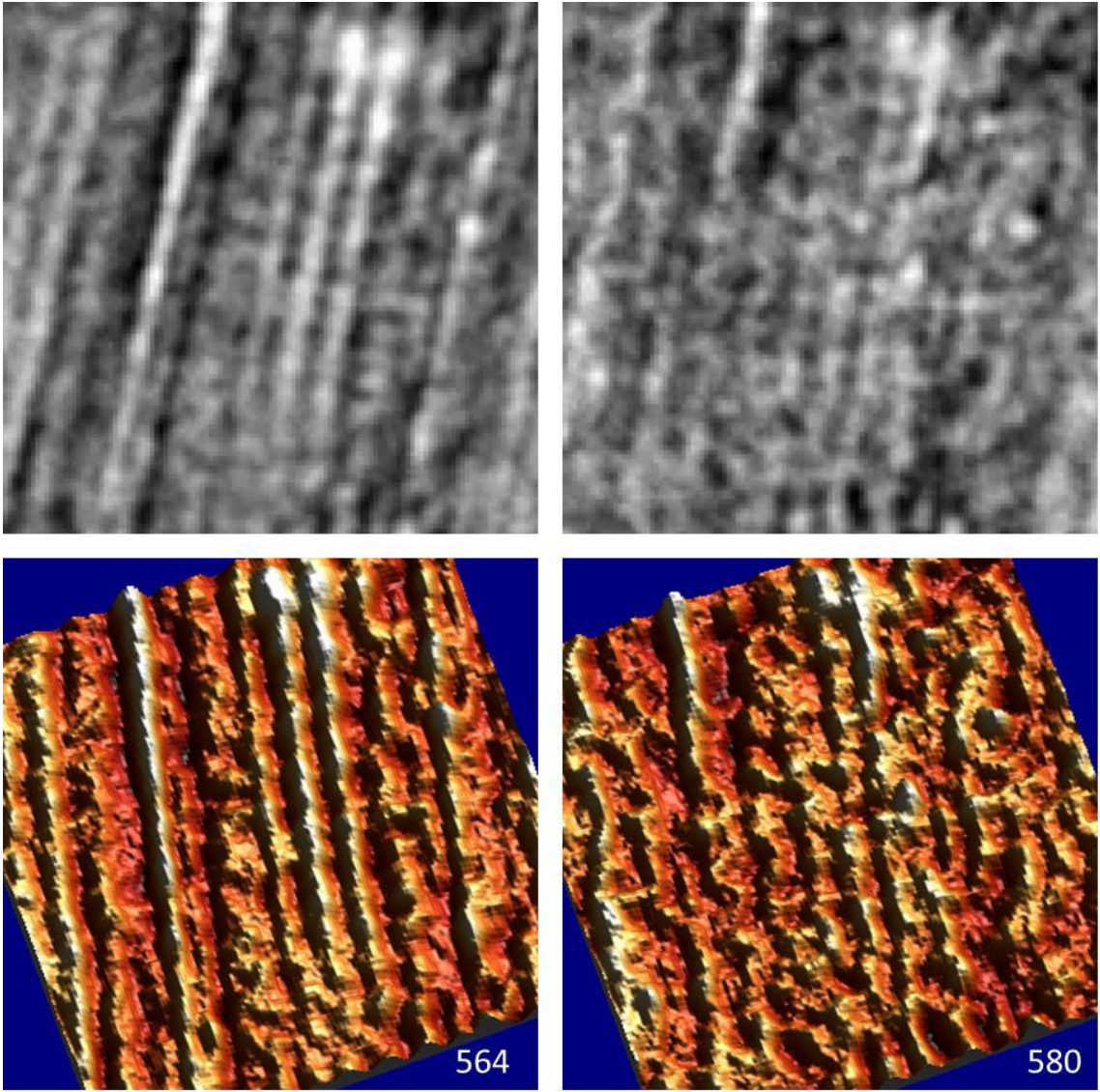

**Fig. 8**

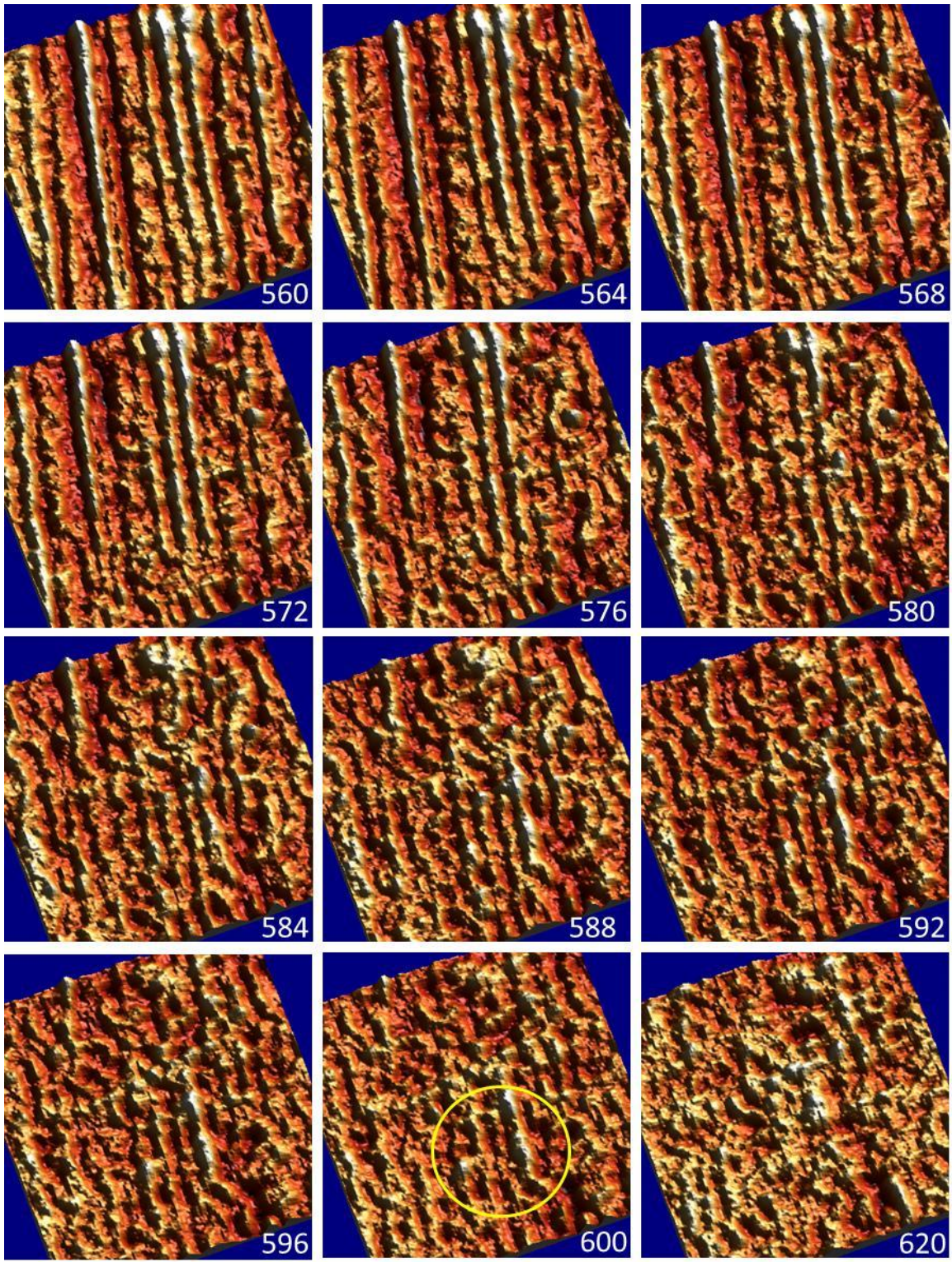

**Fig. 9**

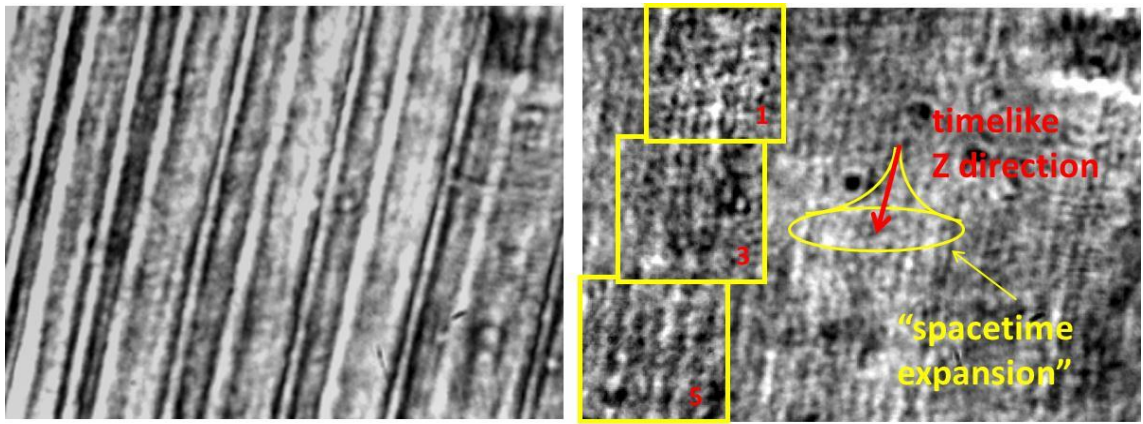

(a)                                    (b)

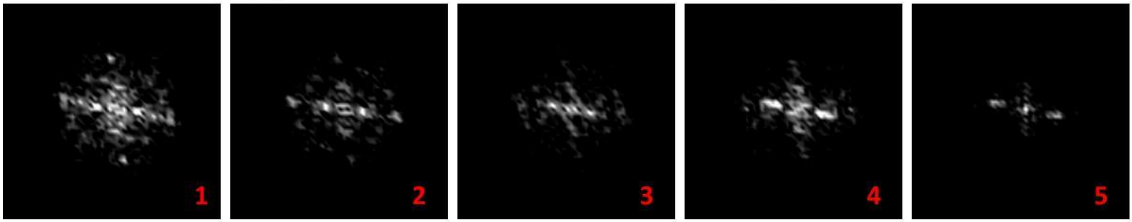

(c)

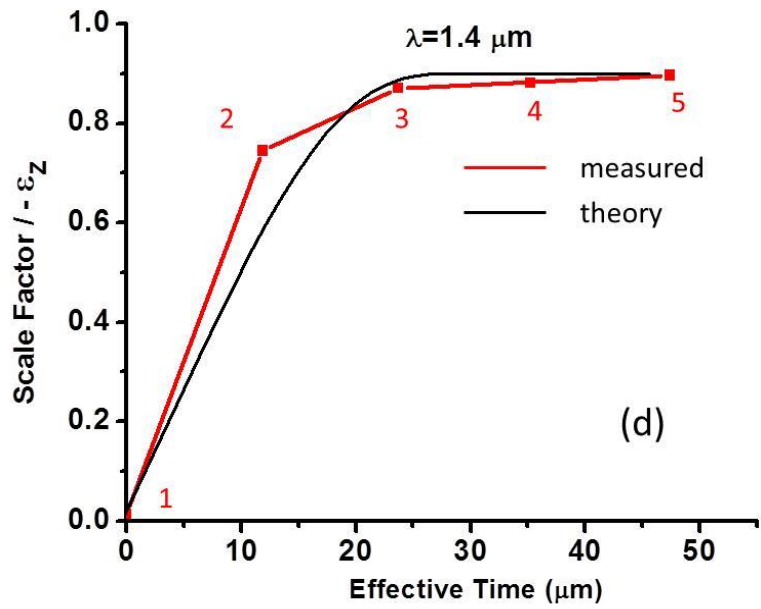

Fig. 10